\documentclass[prb,preprint,superscriptaddress,showpacs]{revtex4}
\usepackage{revsymb}
\usepackage[dvips]{graphicx}
\usepackage{color}
\usepackage{here}
\usepackage{amsfonts}

\begin{document}

\title{ {\LARGE Coupled Chaotic Oscillators and Its Relation with 
Artificial Quadrupeds' Central Pattern Generator} } 

\author{ {\large {\bf Horacio Castellini}} }
\email{hcaste@fceia.unr.edu.ar}
\affiliation{{\rm Dpto\@. de F\'{\i}sica, F\@.C\@.E\@.I\@.A\@., 
Pellegini 250,
2000 Rosario}}

\author{{\large {\bf Hilda Cerderira}}}
\affiliation{{\rm International Centre of Theoretical Physics (ICTP),
Trieste, Italia}}

\author{{\large {\bf Lilia Romanelli}}}
\email{lili@ungs.edu.ar} 
\affiliation{{\rm Instituto de 
Ciencias, Universidad de General
Sarmiento, Roca 850, 1663 San Miguel,
Argentina}}

\begin{abstract}
Animal locomotion employs different periodic patterns known as animal gaits. In
1993 Collins and Stewart achieved the characterization in  quadrupeds and
bipeds by using permutation symmetries groups which impose constrains in the
locomotion centre called Central Generator Pattern (CGP) in the animal brain.
They modelled the CGP by coupling four non linear oscillators and with the only
change in the coupling it is possible to reproduce all the gaits.
In this work we propose to use coupled chaotic oscillators synchronized with
the Pyragas method not only to characterize the CGP symmetries but also
evaluate the time serie behaviour when the foot is in contact with the ground
for futures robotic applications.
\end{abstract}
\maketitle

\section{Introduction}
Lately the analysis of animal gaits have regain interest in between scientist 
of 
different areas. The moderm analysis represents the gait as cyclic patterns in
the movements of symmetrically placed limbs. A cycle means the interval between
successive footstrikes of the same foot during the dynamical process. 
The factor
of footstrike $\beta$ is the fraction of cycle when the foot is in contact with
the surface. 
For simplicity we assume $\beta$ to be the same for all animal feet.
The relative phase of foot ($\phi$) is defined as the fraction of cycle between
the contact of the surface of the foot of reference and another foot when 
this one
enters in contact with surface\cite{r1}. 
In this study the relative phase plays a crucial role to formulate the 
symmetries 
while this is not same the factor of footstrike $\beta$, which will no be taken
into account in this work. 
The mammal phenotypes have evolved in two kind of gaits,
bipedal gaits where the limbs can be out of phase (walking or running) or
in phase (jumping or hopping). 
Quadrupedal gaits with a more complex behaviour of
the realtive phase. The natural gaits are\cite{r2}: 
{\bf Walk}, the limbs move with a quarter
of cycle out of phase, where there is a quarter cycle phase 
difference in between
both fore limbs as well as in between hind lims, 
and half a cycle between diagonals.
{\bf Trot}, the diagonal legs move in phase and this pair is half a 
cycle out of 
phase of the other one. 
{\bf Pace or Rack}, the fore and hind left or roght limbs
are paired and moved half cycle out of phase in between the pairs. 
{\bf Canter},
the right front leg and the left hind leg move in phase, the left front leg and
the right hind leg move half a cycle out of phase in between themselves and out
of phase with respect to the first pair (it was found in horse that pattern
changes from walk to trot, to canter to gallop, when the speed increase). 
{\bf Bound}, the fore legs move in phase, as well as the hind legs, while they
move half a cycle out of pahse. 
{\bf Transverse Gallop}, the front and hind left
(right) legs move one quarter of cycle out of phase, the front (hind) limbs are
slightly out of phase between themselves. {\bf Rotatory Gallop}, similar to
the tranverse gallop but the left and right limbs have their pattern exchange.
{\bf Pronk}, the four limbs move together an in phase.

Biological model assume that the animal nrevous system contains a variety of 
{\em Central Pattern Generators}\cite{r3} (CPG), each oriented to 
specific action. 
For instance, the locomotion CPG controls the rhythm of mammal 
gait\cite{r4}, in
the case of quadrupedal mammals this is modeled by a system of coupled
cell where erach cell is composed by a set of neurons directly responsible to
harmonize the movement of the leg\cite{r5}. 
A simplifield mathematical model of 
locomotion CPG consists in replacing each cell by a non linear 
oscilator\cite{r6}. 
This
model has been studied using differnt method: bifurcation theory, numerical
simulations and phase reponse\cite{r7,r8,r9,r10}.

The idea to study rhythmic patterns in animal gait using symmetries models was
introduced by Hildebrand\cite{r9}, Schoner {\em et al}\cite{r11}. 
The concept of symmetries in 
coupled cells as a model for locomotion CPG in quadrupedal mammals was firt
used by Collins and Stewart\cite{r6}. 
A model for locomotion CPG for quadrupedal mammals
consits in a ring of four coupled nonlinear oscillators. Each oscillator
represents a limb of the animal. The stability and breakdown of the symmetries
play an effective role in the validity of the model. 
Golubitsky {\em et al}\cite{r14}\@. 
argued that symmetries present in above model for walk, trot and pace are not
adequade for quadrupeds, sience the trot and pace correspnd to 
conjugate solutions
which have same stability and they depend on initial conditions. 
Many quadrupeds
move with pace but do not trot (camels) or viceversa (horse), unless they are
trained. In this work we propose a different coupling mechanism in order to
avoid the problem of multiple conjugated solutions\cite{r6}.

The CPG is modeled by the following system of ordinary differential equations:

\begin{equation}
\frac{d X_j}{d t} = f(X_j) + h_j (X_{j-1}, \, X_{j+1})
\label{eq:1}
\end{equation}

where $j=1 \ldots 4$ mod 4 is the index of cell, $X \in \mathbb{R}^n$ is the 
state vector and $f:\mathbb{R}^n \to \mathbb{R}^n$ is a nonlinear velocity 
vector field. We defined the symmetry of ring to the permutations of cell that
preserves the coupling, that is to say a permutation $\sigma$ of 
$\{ 1 \ldots 4 \}$ numbers on the phase space $X=(X_1, X_2, X_3, X_4)$ is:

\begin{equation}
\sigma X=(X_{\sigma^{-1}(1)}, X_{\sigma^{-1}(2)}, X_{\sigma^{-1}(3)}, 
X_{\sigma^{-1}(4)} )
\label{eq:2}
\end{equation}

then $\sigma$ is a symetry of ring in 

\begin{equation}
F(\sigma X)=\sigma F(X).
\label{eq:3}
\end{equation}

where $F(.)=f(.)+h(.)$. Then we deduced the coupling conditions they are:

\begin{equation}
h_j(\sigma X_{j-1},\sigma X_{j+1})=h_{\sigma(j)}(X_{j-1},X_{j+1}).
\label{eq:4}
\end{equation}

If we defined $[i,j]$ as the action to exchange $X_i$ for $X_j$ the symmetries
of the ring are: \{ [1,4]; [2,3]; [1,2]; [4,2] \}. 
From eq-\ref{eq:4} we deduced $h_1(.)=h_3(.)$ and $h_2(.)=h_4(.)$.
These symmetries are called {\bf Type-2} by Collins and Stewart\cite{r6}. 
Another kind of symetry is called 
{\em symmetry of phase change}\cite{r14}. 
Assuming that $X(t)$ is a periodic solution with
minimal period (cycle) {\bf T}, and $\gamma$ represents the symmetry $(ij)$ to
permute $X_i$ for $X_j$, then $\gamma \, X(t)$ will be a periodic solution if
the trajectories $\{X(t)\}_t$ and $\{\gamma X(t) \}_t$ coincide. 
Therefore the only
solution is the existence of pahse slip $\theta$ such that 
$\gamma X_j(t) = X_j(t+\theta)$. The pair $(\gamma,\theta)$ is a 
spatio-temporal
symmetry where $\theta$ is a {\em phase slip}. Finally we define 
as primary gait\cite{r14}, 
those gaits modeled by identical output signal of each cell but out of phase.

We associated the index of cell to each limb as follow, $j=1$ hind left, $j=2$
fore left, $j=3$ fore right and $j=4$ hind right. The possible symmetries of
the primary gait for four legged animals characterized by Type-2 arrays are:

\begin{center}
{\bf Table 1: } Simmetries asocied with gait
\end{center}
\begin{table}[h]
\begin{center}
\begin{tabular}{|l|l|l|}
\hline
Gait & Symmetry & Group \\
\hline
Stopped & $(I,\theta)$ $(\alpha,\theta)$ $(\beta,\theta)$ $(\alpha \beta,\theta)$ &
$D_2 \times S^1$ \\
Pronk & $(I,0)$ $(\alpha,0)$ $(\beta,0)$ $(\alpha \beta,0)$ & $D_2$ \\
Pace & $(I,0)$ $(\alpha,\frac{1}{2})$ $(\beta,\frac{1}{2})$ $(\alpha \beta,0)$ 
& $\tilde{D}_2^D$ \\
Bound & $(I,0)$ $(\alpha,0)$ $(\beta,\frac{1}{2})$ $(\alpha \beta,\frac{1}{2})$  
& $\tilde{D}_2^F$ \\
Trote & $(I,0)$ $(\alpha,\frac{1}{2})$ $(\beta,0)$ $(\alpha \beta,\frac{1}{2})$ 
& $\tilde{D}_2^L$ \\
Rotatory Gallop & $(I,0)$ $(\beta,\frac{1}{2})$  & $\tilde{Z}_2^L$ \\
Transverse Gallop & $(I,0)$ $(\alpha \beta,\frac{1}{2})$ & $\tilde{Z}_2^F$ \\
Canter & $(I,0)$ & $\mathbb{I}$ \\
\hline
\end{tabular}
\label{ta:1}
\end{center}
\caption{
Where $\alpha=(12)(34)$, $\beta=(13)(24)$, $\alpha \beta=(14)(23)$ and $S^1$ refers
to all cyclic group of pahse slip mod 1. $D$ represents the diedral subgroup and
$Z$ all the cyclic subgroups. The tilde indicates the existence of phase slip
symmetry. The notation $\frac{1}{2}$ represents a half cycle out of phase. 
}
\end{table}

Here we study the possibility to use a ring of coupled chaotic oscillators 
to produce
the locomotion of quadrupeds based on the symmetries of primary gaits using 
Pyragas control theory.

\section{A CPG Model}

Here we represented each cell by R\"osler oscillators (see eq.-\ref{eq:5}) coupled 
using Pyragas\cite{r15} method 
with random initial cnditions. We use Rossler chaotic oscillator since this is 
the only one that have shown to synchronize to simulate the primary gait. 
This behaviour does not happen for Van der Pool\cite{r6} or 
Showalter\cite{r16} oscilators even if the
single oscillator reproduce the necessary output. Primary gate is important for any
digital aplication on mechanical limb.

\begin{equation}
\left \{ \begin{array}{rcl}
\frac{d x}{d t} & = & -(y+x)\\
\frac{d y}{d t} & = & x+ 0.2 \, y\\
\frac{d z}{d t} & = & 0.2 \, + \, z \, (x-c)
\end{array} \right.
\label{eq:5}
\end{equation}

We use a direct synchronization mechanism where the master variable is ``$y$'' and 
other one are the slave varibles. We use a delay time serie for obtain 
delay feedback value. Then coupling functions are:

\begin{equation}
\begin{array}{ll}
h_i(X_i,X_{i+1},X_{i-1})= k_i \, (y_{i-1}(t-\tau)-y_i(t)) & \\
+ g_i \, (y_{i+1}(t-\tau)-y_i(t)) & 
\end{array}
\label{eq:6}
\end{equation}

\begin{figure}[H]
\begin{center}
\includegraphics[width=7cm, height=6.5cm]{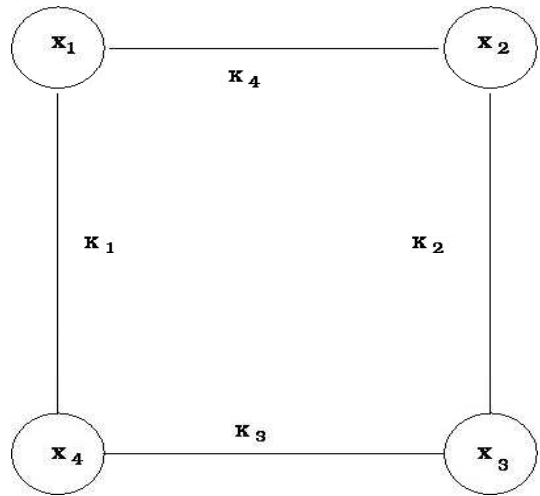}
\end{center}
\caption{Cell coupling diagram.}
\label{fig:1}
\end{figure}

The symmetry conditions associated to a Type-2 array limits the range of $k_i$ and
$g_i$ values. In this case $g_1=k_2$, $g_2=k_3$, $g_3=k_4$, $g_4=k_1$,
$k_1=k_2$ and $k_4=k_3$. An esquematic form it is depected in figure-\ref{fig:1}.
The delay time $\tau \in \mathbb{N}$, and the nonlinear constan $c$ 
(see eq.-\ref{eq:5}) play an important
role in the wave pattern obtained. We consider as output of each cell (oscilator)
the value of the variable $x_i(t)$, where it will be compouse by threshold function:

\begin{equation} 
Q(x)=\left\{ \begin{array}{ll}
0 & \textrm{si $x>2.0$} \\
1 & \textrm{si $x \le 2.0$}
\end{array} \right.
\label{eq:7}
\end{equation}

This defined a mapping from phase space into binary matrix space of 2x2. 
We associate
the value ``1'' to the state ``{\em limb on ground}'' and the value ``0'' to
the state ``{\em limb in movement}'', not on the ground. Finally the matrix 
representation is from now on:

\[
\mathbf{C}= \left( \begin{array}{cc}
\textrm{Fore Left} & \textrm{Fore Right}\\
\textrm{Hind Left} & \textrm{Hind Right}
\end{array} \right)
\]

Therefore the gait is nothing but the sequence of matrices of succesive states 
representing the symmetry of CPG. For instance the pronk is give by the sequence 
of matrices:

\[
\left\{
\left( \begin{array}{cc}
1 & 1 \\
1 & 1 
\end{array} \right) \, ; \,
\left( \begin{array}{cc}
0 & 0 \\
0 & 0 
\end{array} \right) 
\right\}
\]

This allows a clearer visualization of the symmetries of primary gait. Although we
lose the time interval between different patterns. This is highly important for
application in robotics, not only it conditions actuator answers, also for the
fact that mechanical inner may introduce undesirable instabilities. This is
the reason that we have to analyse the time each pattern stays in periodic 
sequence. 

We consider two type of {\em ad hoc}
combinations for the coupling constants, which are the most
represntatives in between the values tried. We call {\bf SA} model when
 $k_1=k_3=0.1$ and $k_2=k_4=0.001$, {\bf SB} a $k_1=k_3=0.1$ and $k_2=k_4=-0.001$.
Those values were selected under the assumption of strong coordination between
the limbs associated to each cerebral hemisphere, while they are weakly 
correlated when they belong to different hemisphere.\footnote{ Remember as
we depected the locomotor CPG by four limb animals in previous paragraphs}

\vspace{1.5cm}
\vspace{0.5cm}

\section{Numerical Results}

\begin{figure}[H]
\begin{center}
\includegraphics[width=8cm, height=8.5cm, angle=-90]{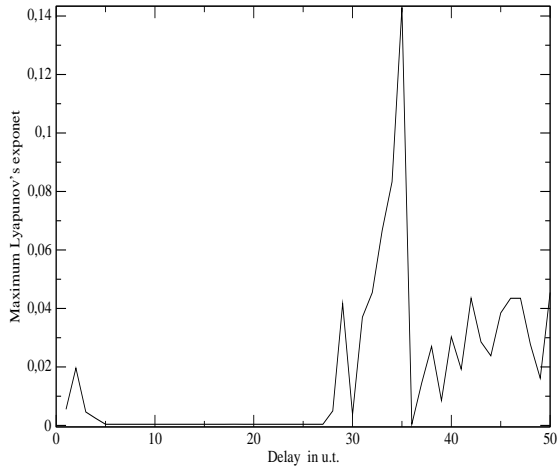}
\end{center}
\caption{Maximum Lyapunov exponent verus delay.}
\label{fig:2}
\end{figure}

There is a dependence of the output, 
on the time delay $\tau$, which is quiet robust
under variations of the parameter $c$ (see eq-\ref{eq:5} ), 
and it is independent of the model
used, {\bf SA} or {\bf SB}. There is a change of state as a function of the time
delay, which follow the order Chaos $\to$ Periodic Oscillations $\to$ Stable Fixed
Point $\to$ Primary Hopf Bifurcation $\to$ Secondary Hopf Bifurcations $to$ Chaos
when $\tau$ increases from zero. This can seen in Fig-\ref{fig:2}, 
where we plot the
maximum Lyapunov exponet of the systems versus $\tau$. 
Since it is necessary to have
a stable time interval in between patterns, we restrict ourselves to the interval
$6 \le \tau \le 38$.

\vspace{1.5cm}
\vspace{0.5cm}

\subsection{SA Coupling}

For $6 \le \tau \le 13$, the periodic gait obtained is:

\[
\left\{
\left( \begin{array}{cc}
1 & 1 \\
1 & 1 
\end{array} \right) \, ; \,
\left( \begin{array}{cc}
0 & 0 \\
0 & 0 
\end{array} \right) 
\right\}
\]

which corresponds to the pronk and the symmetry is $D_2$. On the other hand for
$\tau=14$ the limit cycle is not stable any longer ant it appears an asymptotic
stationary state, which produces the single sequence correspond to the symmetry
$D_2 \times S^1$, therefore a stop, since all limbs are on ground. For $\tau=34$,
the fixed point loses its stability and it becomes unstable, where the orbits 
converge to a single limit cycle. The pattern found for the delay 
$34 \le \tau \le 38$ is :

\[
\left\{
\left( \begin{array}{cc}
0 & 1 \\
1 & 0
\end{array} \right) \, ; \,
\left( \begin{array}{cc}
1 & 1 \\
1 & 1 
\end{array} \right) \, ; \,
\left( \begin{array}{cc}
1 & 0 \\
0 & 1
\end{array} \right) \, ; \,
\left( \begin{array}{cc}
1 & 1 \\
1 & 1 
\end{array} \right)
\right\}
\]

This has a symmetry $\tilde{D}_2^L$, which corresponds to the gait trot. In this
case each step involves the movement of all the limbs, gait which is only observed
in havy quadruopeds, above 1 ton, such as girafes and buffalos\cite{r6}. 
We could not find any other patterns in this interval of time delay.

\subsection{SB Coupling}

For $6 \le \tau \le 13$ the coupled chaotic system oscillate in a stable 
limit cycle,
which in this case produce the patterns:

\[
\left\{
\left( \begin{array}{cc}
1 & 1 \\
0 & 0
\end{array} \right) \, ; \,
\left( \begin{array}{cc}
1 & 1 \\
1 & 1 
\end{array} \right) \, ; \,
\left( \begin{array}{cc}
0 & 0 \\
1 & 1
\end{array} \right) \, ; \,
\left( \begin{array}{cc}
1 & 1 \\
1 & 1 
\end{array} \right) 
\right\}
\]

which has symmetry $\tilde{D}_2^F$ and corresponds to a gait bound. We should
mention that the true bound as the one observed in 
siberian squirrel\cite{r6} implies
the state of all limbs on air. On the other hand in this case all four limbs
are on the ground in order to generate another step which does not exist in
nature. This drawback, we resolve applying ``{\em not}'' operator each matrix.
As in the {\bf SA} case, for $14 \le \tau \le 32$ the coupled system has 
stable fixed point and the limit cycle becomes unstable, which correspond
to stop. For $\tau=33$ the fix point become unstable and a stable cycle appears
producing the patterns:

\[
\left\{
\left( \begin{array}{cc}
1 & 1 \\
1 & 1
\end{array} \right) \, ; \,
\left( \begin{array}{cc}
1 & 0 \\
1 & 0 
\end{array} \right) \, ; \,
\left( \begin{array}{cc}
1 & 1 \\
1 & 1
\end{array} \right) \, ; \,
\left( \begin{array}{cc}
0 & 1 \\
0 & 1 
\end{array} \right)
\right\}
\]

which corresponds to gait pace with symmetry $\tilde{D}_2^D$. 
As describes above the
horse never sets all four limbs on the ground, then applying ``{\em not}'' 
operator each matrix, we have resolved it. For this coupling the gait is not 
structurally stable for all time delay values. From $\tau=34$ until 4 $\tau=37$
the gait changes to another periodic pattern:

\[
\left\{
\left( \begin{array}{cc}
0 & 1 \\
1 & 1
\end{array} \right) \, ; \,
\left( \begin{array}{cc}
1 & 1 \\
1 & 0 
\end{array} \right) \, ; \,
\left( \begin{array}{cc}
1 & 0 \\
1 & 0
\end{array} \right) \, ; \,
\right.
\]
\[
\left.
\left( \begin{array}{cc}
1 & 0 \\
1 & 1 
\end{array} \right) \, ; \,
\left( \begin{array}{cc}
1 & 1 \\
0 & 1 
\end{array} \right) \, ; \,
\left( \begin{array}{cc}
0 & 1 \\
0 & 1 
\end{array} \right)
\right\}
\]

With symmetry $\{ (I,0) \, (\alpha,\frac{2}{3}) \, (\beta,\frac{2}{3}) \}$.
This symmetry does not correspond to any primary gait observed in nature.
For $\tau=38$ the system generates again a gaite like pace.

\section{Conclusions}

Pyraga's direct synchronization (see eq \ref{eq:6}) is a novel coupling mechanism
between cell for Type-2 networks when it is used as CPG model. Within this 
we can avoid the conjugate undesirable solutions. But no natural patterns appear 
in {\bf SB} model. Also we can not look for {\em canter} and 
{\em transverse gallop} gaits. This network type is not adequately for a 
natural GCP model. However it is useful as an artificial CPG mechanism in
robotic science.

\section{References}

\begin{thebibliography}{14}
\expandafter\ifx\csname natexlab\endcsname\relax\def\natexlab#1{#1}\fi
\expandafter\ifx\csname bibnamefont\endcsname\relax
  \def\bibnamefont#1{#1}\fi
\expandafter\ifx\csname bibfnamefont\endcsname\relax
  \def\bibfnamefont#1{#1}\fi
\expandafter\ifx\csname citenamefont\endcsname\relax
  \def\citenamefont#1{#1}\fi
\expandafter\ifx\csname url\endcsname\relax
  \def\url#1{\texttt{#1}}\fi
\expandafter\ifx\csname urlprefix\endcsname\relax\def\urlprefix{URL }\fi
\providecommand{\bibinfo}[2]{#2}
\providecommand{\eprint}[2][]{\url{#2}}

\bibitem[{\citenamefont{Gray}(1968)}]{r1}
\bibinfo{author}{\bibfnamefont{J.}~\bibnamefont{Gray}},
  \emph{\bibinfo{title}{Animal Locomotion}} (\bibinfo{publisher}{Weidenfeld and
  Nicolson}, \bibinfo{year}{1968}).

\bibitem[{\citenamefont{Alexander}(1984)}]{r2}
\bibinfo{author}{\bibfnamefont{R.}~\bibnamefont{Alexander}},
  \bibinfo{journal}{Int. J. Robot Res.} \textbf{\bibinfo{volume}{3}},
  \bibinfo{pages}{49} (\bibinfo{year}{1984}).

\bibitem[{\citenamefont{A.~Cohen and Grillner}(1988)}]{r3}
\bibinfo{author}{\bibfnamefont{S.~R.} \bibnamefont{A.~Cohen}} \bibnamefont{and}
  \bibinfo{author}{\bibfnamefont{S.}~\bibnamefont{Grillner}},
  \emph{\bibinfo{title}{Neural Control of Rhythmic Movements in Vertebrates}}
  (\bibinfo{publisher}{Wiley New York}, \bibinfo{year}{1988}).

\bibitem[{\citenamefont{Dagg}(1973)}]{r4}
\bibinfo{author}{\bibfnamefont{A.}~\bibnamefont{Dagg}},
  \bibinfo{journal}{Mammal Rev.} \textbf{\bibinfo{volume}{3}},
  \bibinfo{pages}{135} (\bibinfo{year}{1973}).

\bibitem[{\citenamefont{Glass and Young}(1979)}]{r5}
\bibinfo{author}{\bibfnamefont{L.}~\bibnamefont{Glass}} \bibnamefont{and}
  \bibinfo{author}{\bibfnamefont{R.}~\bibnamefont{Young}},
  \bibinfo{journal}{Brain Res.} \textbf{\bibinfo{volume}{179}},
  \bibinfo{pages}{207} (\bibinfo{year}{1979}).

\bibitem[{\citenamefont{Collins and Stewart}(1993)}]{r6}
\bibinfo{author}{\bibfnamefont{J.}~\bibnamefont{Collins}} \bibnamefont{and}
  \bibinfo{author}{\bibfnamefont{I.}~\bibnamefont{Stewart}},
  \bibinfo{journal}{J. Nonlin. Sci.} \textbf{\bibinfo{volume}{3}},
  \bibinfo{pages}{345} (\bibinfo{year}{1993}).

\bibitem[{\citenamefont{Alexander and Goldspink}(1977)}]{r7}
\bibinfo{author}{\bibfnamefont{R.}~\bibnamefont{Alexander}} \bibnamefont{and}
  \bibinfo{author}{\bibfnamefont{J.}~\bibnamefont{Goldspink}},
  \emph{\bibinfo{title}{Mechanics and Energetics of Animal Locomotion}}
  (\bibinfo{publisher}{Chapman and Hall}, \bibinfo{year}{1977}).

\bibitem[{\citenamefont{Hildebrand}(1964)}]{r8}
\bibinfo{author}{\bibfnamefont{M.}~\bibnamefont{Hildebrand}},
  \bibinfo{journal}{Folia Biotheoretica} \textbf{\bibinfo{volume}{4}},
  \bibinfo{pages}{10} (\bibinfo{year}{1964}).

\bibitem[{\citenamefont{Hildelbrand}(1965)}]{r9}
\bibinfo{author}{\bibfnamefont{M.}~\bibnamefont{Hildelbrand}},
  \bibinfo{journal}{Science} \textbf{\bibinfo{volume}{150}},
  \bibinfo{pages}{701} (\bibinfo{year}{1965}).

\bibitem[{\citenamefont{Collins and Richarmond}(1994)}]{r10}
\bibinfo{author}{\bibfnamefont{J.}~\bibnamefont{Collins}} \bibnamefont{and}
  \bibinfo{author}{\bibfnamefont{S.}~\bibnamefont{Richarmond}},
  \bibinfo{journal}{Biol. Cybern.} \textbf{\bibinfo{volume}{71}},
  \bibinfo{pages}{375} (\bibinfo{year}{1994}).

\bibitem[{\citenamefont{C.~Canvier and Byrne}(1997)}]{r11}
\bibinfo{author}{\bibfnamefont{C.~B.} \bibnamefont{C.~Canvier}}
  \bibnamefont{and} \bibinfo{author}{\bibfnamefont{J.}~\bibnamefont{Byrne}},
  \bibinfo{journal}{Biol. Cybern.} \textbf{\bibinfo{volume}{68}},
  \bibinfo{pages}{1} (\bibinfo{year}{1997}).

\bibitem[{\citenamefont{Golubistsky and Luciano}(2001)}]{r14}
\bibinfo{author}{\bibfnamefont{M.}~\bibnamefont{Golubistsky}} \bibnamefont{and}
  \bibinfo{author}{\bibfnamefont{P.}~\bibnamefont{Luciano}},
  \bibinfo{journal}{J. Math. Biol.} \textbf{\bibinfo{volume}{42}},
  \bibinfo{pages}{291} (\bibinfo{year}{2001}).

\bibitem[{\citenamefont{Pyragas}(1992)}]{r15}
\bibinfo{author}{\bibfnamefont{K.}~\bibnamefont{Pyragas}},
  \bibinfo{journal}{Phy. Lett.} \textbf{\bibinfo{volume}{170}}
  (\bibinfo{year}{1992}).

\bibitem[{\citenamefont{V.~Petrov and Showalter}(1992)}]{r16}
\bibinfo{author}{\bibfnamefont{K.~S.} \bibnamefont{V.~Petrov}}
  \bibnamefont{and}
  \bibinfo{author}{\bibfnamefont{K.}~\bibnamefont{Showalter}},
  \bibinfo{journal}{J. Chem. Phys.} \textbf{\bibinfo{volume}{97}},
  \bibinfo{pages}{6191} (\bibinfo{year}{1992}).

\end{thebibliography}

\end{document}